\documentstyle[12pt]{article}
\title{
{\vspace{-3cm} \normalsize \hfill MS-TPI-92-13
                                            }\\[25mm]
Numerical Investigation of the
Interface Tension in the three-dimensional Ising Model}

\author{Sabine Klessinger and Gernot M\"unster \\
        Institut f\"ur Theoretische Physik I,
        Universit\"at M\"unster\\
        Wilhelm-Klemm-Str.~9, D-4400 M\"unster, Germany}
\date{May 25, 1992}

\newcommand{\be}{\begin{equation}}
\newcommand{\ee}{\end{equation}}

\begin{document}
\maketitle

\begin{abstract}
The interface tension in the three-dimensional Ising model in the low
temperature phase is investigated by means of the Monte Carlo method.
Together with other physically relevant quantities it is obtained from a
calculation of time-slice correlation functions in a cylindrical
geometry.
The results at three different values of the temperature are compared
with the predictions from a semiclassical approximation in the framework
of renormalized $\phi^4$ theory in three dimensions, and are in good
agreement with them.
\end{abstract}
\newpage
\section{Introduction}
In various systems in statistical mechanics or in nature at
temperatures $T$ below a certain critical temperature $T_c$ interfaces
(domain walls) may be present, separating coexisting phases.
The interface tension $\tau$ is the free energy per unit area of
interfaces.
It is an interesting quantity for various reasons.
For numerical simulations of field theories in a phase with broken
discrete symmetry finite volume effects are often dominated by tunneling
\cite{AHK2}.
The size of these finite volume effects is intimately related to the
interface tension.
Secondly, the interface tension is relevant for first order phase
transitions, where it determines the nucleation rate.

For many binary systems the interface tension $\tau$ has been
investigated experimentally as a function of the temperature $T$
\cite{Pegg,Gielen,Moldover,Chaar}.
As $T$ increases towards $T_c$ the reduced interface tension
\be
\sigma = \frac{\tau}{kT} \,,
\ee
where $k$ is Boltzmann's constant, vanishes according to the scaling law
\be
\sigma \sim \sigma_0 t^{\mu} \,,
\ee
where
\be
t = \left|\frac{T-T_c}{T_c}\right| \,,
\ee
and $\sigma_0$ is the critical amplitude of the interface tension.
Widom's scaling law \cite{Widom2,Widom1},
\be
\mu = 2\nu \,,
\ee
relates the universal critical exponent $\mu$ to the critical exponent
of the correlation length $\xi$:
\be
\xi \sim \xi_0 t^{-\nu} \,.
\ee
The experimental results \cite{Widom1,Pegg,Gielen,Moldover} for the
critical index $\mu$,
\be
\mu = 1.26 \pm 0.01 \,,
\ee
and for the critical exponent $\nu$ \cite{Beysens,Sengers,Kumar}, as
well as the prediction from the renormalization group \cite{Zinn},
\be
\nu = 0.630 \pm 0.002 \,,
\ee
are consistent with Widom's law.

Other quantities, which are universal according to the scaling
hypothesis \cite{Fisk,Stauffer}, are the dimensionless products of
critical amplitudes
\be
R_+ = \sigma_0 (\xi_0^+)^2 \,, \hspace{1cm}
R_- = \sigma_0 (\xi_0^-)^2 \,,
\ee
where $\xi_0^+$ and $\xi_0^-$ are the critical correlation length
amplitudes in the high-temperature and low-temperature phases,
respectively.

On the theoretical side we dispose of two models in the relevant
universality class, the Ising model and the $\phi^4$ field theory in
three Euclidean dimensions.
Early field theoretic calculations \cite{Pant,BF} of $R_-$ (and $R_+$)
in the framework of the $\epsilon$-expansion, where $d=4-\epsilon$ is
the number of dimensions of space, were in disagreement with the
experimental results.
The discrepancy could, however, be resolved by a calculation in
renormalized field theory directly in three dimensions \cite{Mue3d}.
This calculation was performed by means of a semiclassical saddle-point
expansion to one-loop order around the kink-solution.
The semiclassical method yields an expression for the energy splitting
between the two ``vacua'' due to tunneling in a finite volume, which
involves the interface tension.
For some quantities related to the energy splitting and the interface
tension it yields definite predictions in terms of the renormalized
parameters of the model (see below).

A similar development took place in the case of Monte Carlo simulations
of the Ising model.
First results for the interface tension \cite{Binder}, which were at
variance with the (real) experiments, have been improved by recent
numerical studies \cite{Mon,Meyer}, which made the discrepancy vanish.

A systematic numerical study of the interface tension and related
objects, which would allow a comparison with the semiclassical
approximation, was, however, completed only recently for the
four-dimensional Ising model \cite{JanShen}.

In this article we present the results of a Monte Carlo investigation of
the three-dimensional Ising model.
The interface tension is determined from the volume dependence of the
energy splitting, a method that was introduced in \cite{AHK1}.
Furthermore, masses and renormalized parameters are calculated, which
are necessary for a comparison with the semiclassical formulae.

Let us add a remark about a controversy, which stems from work of Borgs
and Imbrie \cite{BoIm}.
They derive a formula for the asymptotic large volume behaviour of the
energy splitting, which disagrees with the semiclassical result.
Their result, however, is valid only at very low temperatures, where
interfaces are rigid.
In three dimensions this low temperature region is separated from
the scaling region in the low temperature phase by the roughening
transition \cite{Rough}.
Above the roughening transition there are massless interface modes and
the results of \cite{BoIm} do not apply to the rough phase.

In four dimensions, however, there is a conflict between \cite{BoIm} and
the semiclassics, because of the absence of a roughening transition.
The Monte Carlo results of \cite{JanShen} are in favour of the
semiclassical formula.
Nevertheless it might be possible that Borgs and Imbrie are right for
very large volumes.
Namely, if one adopts the common assumption that the interface dynamics
is effectively described by a SOS-model or equivalently a discrete
Gaussian model, the results of Mack and G\"opfert \cite{MaGo} suggest
that in the scaling region there is a very small mass in the interface
dynamics, which only shows up in the behaviour at very large volumes.
\section{The three-dimensional Ising model}
We consider the Ising model on a three-dimensional cubic lattice of
cylindrical geometry.
This means that the lattice has a quadratic cross-section of area $L^2$
in two directions and extends over $T$ lattice units in the third
direction, such that $L \ll T$.
($T$ is not to be confused with the physical temperature, which does not
appear any more from here on.)
The Boltzmann factor is denoted
\be
e^{- \beta {\cal H}} \,,
\ee
with
\be
{\cal H} = - \sum_{x} \sum_{\mu=1}^3 \phi_x \phi_{x+\hat{\mu}}
\hspace{1ex} , \hspace{4ex} \beta > 0 \, ,
\ee
where $\hat{\mu}$ denotes the unit vector in the positive
$\mu$-direction.
The variables $\phi_x$ are associated with the lattice points and assume
values $\phi_x = \pm1$.
They obey periodic boundary conditions.
The transfer matrix is denoted $e^{-H}$, where $H$ is the
Hamiltonian in the language of quantum theory.
The lowest eigenvalue of $H$ is normalized to zero.

The three-dimensional Ising model in an infinite volume is known to
possess a critical point at \cite{Adler,Pawl}
\be
\beta_c = 0.221\,654\,(6) \,.
\ee
For $\beta > \beta_c$ the $Z_2$-symmetry $\phi \rightarrow -\phi$
is spontaneously broken and the field acquires a non-zero vacuum
expectation value, e.g.
\be
\langle \phi_x \rangle = \pm v , \hspace{4ex} v > 0 .
\ee
In this phase the spectrum of the transfer matrix is doubly degenerate.
In particular there are two ground states or ``vacua''
$| 0_{\pm} \rangle$ with
\be
\langle 0_+ | \phi(x) | 0_+ \rangle = v \,, \hspace{1cm}
\langle 0_- | \phi(x) | 0_- \rangle = -v \,.
\ee

On the other hand it is well known in statistical mechanics that
spontaneous symmetry breaking does not occur in a finite volume.
Let us consider a situation in which $L$ is finite.
As a consequence of the Frobenius-Perron theorem applied to the transfer
matrix \cite{Domb} there is a unique ground state $| 0_s \rangle $
symmetric under the reflection $\phi \rightarrow -\phi$, and the vacuum
expectation value of the field vanishes.
This means that the degeneracy of the infinite volume ground states
$| 0_{\pm} \rangle $ is lifted.
Separated from the ground state $| 0_s \rangle $ by a small energy
splitting $ E_{0a} $ there is an antisymmetric state $| 0_a \rangle $,
and if one decomposes these states as
\be
| 0_s \rangle \equiv \frac{1}{\sqrt{2}}
\left( | 0_+ \rangle + | 0_- \rangle \right) \hspace{2cm}
| 0_a \rangle \equiv \frac{1}{\sqrt{2}}
\left( | 0_+ \rangle - | 0_- \rangle \right) \,,
\ee
then $| 0_+ \rangle $ and $| 0_- \rangle $ are states which go over into
the degenerate vacua in the infinite volume limit.

The energy splitting $E_{0a}$ is due to tunneling between
$| 0_+ \rangle$ and $| 0_- \rangle $ in a finite volume.
Its volume dependence was studied in \cite{Fisher,PF}.
Their analysis is based on a picture of domains which extend over
the spatial volume and cover certain intervals in time.
Neighbouring domains with a different sign of the field are
separated by domain walls, which can be considered as tunneling events.
{}From this picture a prediction about the energy splitting of the
form
\be
E_{0a} \sim \exp \left\{ -\sigma L^2 \right\}
\ee
is obtained, where $\sigma$ is the interface tension associated with the
domain walls.
One sees that tunneling effects vanish very rapidly with increasing
volume.

The next higher states are the symmetric and antisymmetric one-particle
states with energies $E_{1s}$ and $E_{1a}$, respectively.
In the limit $L \rightarrow \infty$ they become degenerate too, and
their energy defines the physical mass $m$.

In a numerical simulation the low-lying spectrum of zero-momentum
states can be obtained from correlations of time-slices
\be
S_t \equiv \frac{1}{L^2} \sum_{{\bf x}} \phi_{{\bf x},t}
\; , \hspace{1cm} x=({\bf x},t) .
\ee
Let
\be
Z \equiv \mbox{Tr\,} e^{-TH}
= 1 + e^{-TE_{0a}} + e^{-TE_{1s}} + e^{-TE_{1a}} + \cdots
\ee
be the partition function.
Then the vacuum expectation value of the product of timeslice field
averages is given by
$$
\langle S_0 S_t \rangle Z \equiv
\mbox{Tr\,} \left\{ S_0 e^{-tH} S_t e^{-(T-t)H} \right\}
$$
$$
= v^2    \left[ e^{-tE_{0a}} + e^{-(T-t)E_{0a}} \right] +
c_{01}^2 \left[ e^{-tE_{1a}} + e^{-(T-t)E_{1a}} \right]
$$
$$
+ c_{10}^2    \left[ e^{-tE_{0a}-(T-t)E_{1s}} +
                     e^{-(T-t)E_{0a}-tE_{1s}} \right]
$$
\be \label{S0St}
+ c_{11}^2    \left[ e^{-tE_{1a}-(T-t)E_{1s}} +
                     e^{-(T-t)E_{1a}-tE_{1s}} \right] + \cdots
\ee
where the matrix elements are defined as
\be
v \equiv \langle 0_s | S_t | 0_a \rangle
\ee
\be
c_{01} \equiv \langle 0_s | S_t | 1_a \rangle \hspace{1cm}
c_{10} \equiv \langle 1_s | S_t | 0_a \rangle \hspace{1cm}
c_{11} \equiv \langle 1_s | S_t | 1_a \rangle \,.
\ee
The field expectation value $v$ and the energy splitting
$E_{0a}$ can be extracted from the $t$-dependence of
$\langle S_0 S_t \rangle$, if $T$ and $t$ are chosen suitably large.
We define the connected time-slice correlation function in a finite
volume by
\be
\langle S_0 S_t \rangle_c \equiv
\langle S_0 S_t \rangle - \frac{v^2}{Z}
\left( e^{-tE_{0a}} + e^{-(T-t)E_{0a}} \right) \,.
\ee
Here one can use the approximation
\be
Z = 1 + e^{-TE_{0a}}
\ee
for large enough $T$.
The susceptibility is given by
\be
\chi_2 \equiv L^2 \sum_t \langle S_0 S_t \rangle_c \,,
\ee
and the second moment of the correlation function is defined by
\be
\mu_2 \equiv 3 L^2 \sum_t \langle S_0 S_t \rangle_c \,
\sin^2 \frac{\pi t}{T} \left( \sin^2 \frac{\pi}{T} \right)^{-1} \,.
\ee

The correlation function of squared time-slices,
$\langle S_0^2 S_t^2 \rangle$, has an expansion similar to (\ref{S0St}),
from which the energies $E_{1s}$ and $E_{2s}$ can be obtained.

Near the critical point, where one is deep enough in the scaling region,
the critical behaviour can be described in terms of the
three-dimensional $\phi^4$ theory, which is in the same universality
class.
The quantities, which are usually handled in field theory, are defined
differently from those, which are common in statistical physics.
For the definition of the renormalized mass $m_R$, the wavefunction
renormalization factor $Z_R$ and the renormalized field expectation
value $v_R$ we refer to \cite{AHK2}.
The relation to the quantities introduced above is in three
dimensions given by
\be
m_R^2 = \frac{6 \chi_2}{\mu_2} \,,\hspace{1cm}
Z_R = \frac{6 \beta \chi_2^2}{\mu_2} \,,\hspace{1cm}
v_R = \sqrt{\beta / Z_R}\ v \,.
\ee
A suitably defined dimensionless renormalized coupling is
\be
u_R \equiv 3\,\frac{m_R}{v_R^2} \,.
\ee

The energy splitting $E_{0a}$ can be calculated in a semiclassical
approximation, which amounts to a saddle-point expansion around the
classical kink solution of $\phi^4$ theory.
The kink interpolates between the two field values at the minima of the
potential and represents an interface separating regions with different
local mean values of the field.
The calculation including quadratic fluctuations around the kink
solution has been done in \cite{Mue4d} for d=4 dimensions and in
\cite{Mue3d} for the three-dimensional case.
In three dimensions the result is
\be
E_{0a} = C \exp \left\{ -\sigma(L) L^2 \right\} ,
\ee
where $\sigma$ is the interface tension, and
\be \label{C}
C = 4 \,\frac{\Gamma(3/4)}{\Gamma(1/4)} \,\sqrt{\frac{2}{u_R}} \ m \,.
\ee
The interface tension has a negligible exponentially small
$L$-dependence and the value at $L = \infty$ is
\be \label{sigma}
\sigma_{\infty} = 2 \,\frac{m^2}{u_R}
   \left( 1 -
   \frac{u_R}{4 \pi}
   \left( \frac{39}{32} - \frac{15}{16} \log 3 \right)
   + {\cal O}(u_R^2)
   \right) .
\ee
Since $m$ is the inverse of the correlation length in the low
temperature phase, we obtain an expression for the universal amplitude
product
\be
R_- = \frac{2}{u_R^*} \left( 1 - \frac{u_R^*}{4 \pi}
   \left( \frac{39}{32} - \frac{15}{16} \log 3 \right)
   + {\cal O}(u_R^{*2})
   \right) ,
\ee
where $u_R^*$ is the value of $u_R$ at the critical point.
{}From a combination of various analytical results in the literature it
can be estimated to be
\be \label{uR}
u_R^* = 15.1 \pm 1.3 \;.
\ee
which yields
\be \label{Rsemi}
R_- = 0.1024 \pm 0.0088
\ee
and
\be
C /m = 0.49 \pm 0.02
\ee
for $\beta \rightarrow \beta_c$.
\section{Monte Carlo calculation}
We performed Monte Carlo simulations of the three-dimensional Ising
model on lattices with an extension of $T = 120$ in the third
direction.
For the simulation the algorithm of Swendsen and Wang \cite{SWA} was
employed.
Wolff's one-cluster version \cite{Wolff}, which usually performs
better, was not used, since we wanted to take advantage of the improved
estimators for four-point functions.
Three different values of the inverse temperature $\beta$ were taken,
namely $\beta = 0.2275, \ 0.2327$ and $0.2391$.
The size $L$ was varied in steps of 2 in the ranges 10--18, 8--14 and
4--10, respectively.
{}From the correlation functions of time-slices and squared time-slices
the field expectation value $v$ and the energies $E_{0a},\ E_{1a},\
E_{1s}$ and $E_{2s}$ were obtained by means of fits to the
$t$-dependence.
Furthermore the susceptibility $\chi_2$ and the second moment $\mu_2$
have been calculated.
The results for the energies are displayed in table 1.
The table also contains estimates for the physical mass $m$, which are
obtained by weighted averages of $E_{1s}$ and $E_{1a}$ on the largest
available lattices.
For the $\beta$-value closest to the critical point the $L$-dependence
of the low-lying spectrum is shown in figure~1.

For each value of $\beta$ the energy splittings have been fitted
according to
\be \label{E0afit}
E_{0a} = C \exp \left\{ -\sigma L^2 \right\} ,
\ee
in order to get the interface tension and the prefactor $C$.
The fits are displayed in figure 2 and the numerical results are
collected in table 2.

The values for $\sigma$ are then fitted to the scaling law in both forms
\be
\sigma = \sigma_0 |1 - \beta/\beta_c|^{\mu}
\ee
and
\be \label{sigscal}
\sigma = \sigma_0 |1 - \beta_c/\beta|^{\mu} \,.
\ee
The results are
\be
\mu = 1.22(1) \,, \hspace{1cm} \sigma_0 = 1.29(4)
\ee
and
\be
\mu = 1.28(1) \,, \hspace{1cm} \sigma_0 = 1.64(5) \,,
\ee
respectively.
The fit (\ref{sigscal}) is drawn in figure 3.
The corresponding fits for the physical mass, which is equal to the
inverse correlation length $\xi$, yield
\be
\nu = 0.61(5) \,, \hspace{1cm} m_0 = 3.8(6)
\ee
and
\be
\nu = 0.64(6) \,, \hspace{1cm} m_0 = 4.3(7) \,.
\ee
The differences are due to the fact that one is not close enough to the
critical point, but the critical exponents are well consistent with the
known values from the literature.
Forming the universal amplitude ratio we obtain
\be
R_- = 0.09(3) \,,
\ee
in excellent agreement with the estimate (\ref{Rsemi}).

{}From the correlation function of time slices the field theoretic
quantities have been calculated too.
The results are contained in table 3.
The second moment $\mu_2$ has a relatively large error.
This is due to the fact that it is sensitive to the correlation function
at larger distances, where the signal-to-noise ratio is smaller.
As a consequence also the estimates for $m_R$ and the coupling $u_R$ are
afflicted by large uncertainties.
The numerical values for $m_R$ are, however, well consistent with
those for $m$, as expected from series expansion results \cite{TaFi}.
Furthermore the values for $u_R$ agree with the estimate (\ref{uR})
within errors.

In order to compare with the semiclassical formulae we formed the ratios
$\sigma / m^2$ and $C/m$ from the measured values.
The results are shown in table 4.
Although in the temperature range under consideration $\sigma$ and $C$
are varying by factors of 4 and 2, respectively, the dimensionless
ratios are nearly constant, supporting the semiclassical prediction.
The corresponding couplings $u_R$, calculated according to (\ref{sigma})
and (\ref{C}) are also shown.
As the table reveals the couplings obtained in this way are in good
agreement with each other for each value of $\beta$.
They are also consistent with the measured couplings as well as with the
estimate (\ref{uR}).
\section{Conclusion}
The Monte Carlo simulation of the three-dimensional Ising model on a
lattice with cylindrical geometry allowed to determine the energy
splitting $E_{0a}$ in a finite volume with an accuracy sufficient for a
calculation of the interface tension $\sigma$.
The scaling law for $\sigma$ and for the inverse correlation length $m$
was verified and the resulting amplitude product $R_-$ is in very good
agreement with an estimate based on a semiclassical approximation in
the framework of $\phi^4$ theory.
The dependence of $\sigma$ and the prefactor $C$ on the renormalized
mass $m_R$ and on the renormalized dimensionless coupling $u_R$
predicted by the semiclassical approximation to one-loop order is in
agreement with the results of the numerical simulation.

{\bf Acknowledgement}: The calculations have been performed on the
``Landesvektorrechner'' of the RWTH Aachen and on the VAX 9000 of the
University of M\"unster.

\newpage
\begin{center}     \Large\bf Table 1  \rm\normalsize
\end{center}
The low-lying energies on lattices of size $L^2 \cdot T$.
The physical mass $m$ in the last row is the weighted average of
$E_{1s}$ and $E_{1a}$ on the largest available lattices.
\begin{center}
\begin{tabular}{|@{\,}r@{}r@{ }||l|l|l|l|l|} \hline

&&\hspace{20pt}$E_{0a}$&\hspace{10pt}$E_{1s}$&\hspace{12pt}$E_{1a}$
&\hspace{10pt}$E_{2s}$&\hspace{20pt}$m$\\ \hline
\multicolumn{7}{|c|}{\rule[-1.5ex]{0em}{4ex} \bf $\beta = 0.2275$}\\
\hline
L = &10&0.04334(8)&0.42(4)&0.69(6)&1.4(4)&\\ \hline
L =&12&0.0217(1)&0.36(4)&0.55(3)&1.0(2)&\\ \hline
L =&14&0.00989(7)&0.35(4)&0.455(16)&0.84(14)&\\ \hline
L =&16&0.00400(7)&0.38(5)&0.410(13)&1.0(2)&\\ \hline
L =&18&0.00151(8)&0.39(3)&0.420(12)&1.0(4)&0.416(16)\\ \hline
\multicolumn{7}{|c|}{\rule[-1.5ex]{0em}{4ex} \bf $\beta = 0.2327$}\\
\hline
L =&8&0.0348(1)&0.54(4)&0.90(15)&1.4(2)&\\ \hline
L =&10&0.01069(7)&0.51(2)&0.67(2)&1.2(1)&\\ \hline
L =&12&0.00253(10)&0.54(2)&0.606(20)&1.1(2)&\\ \hline
L =&14&0.00047(20)&0.59(7)&0.62(1)&1.5(5)&0.619(14)\\ \hline
\multicolumn{7}{|c|}{\rule[-1.5ex]{0em}{4ex} \bf $\beta = 0.2391$}\\
\hline
L =&4&0.174(1)&1.05(2)&5(3)&1.8(3)&\\ \hline
L =&6&0.0491(2)&0.74(7)&1.35(9)&2.1(5)&\\ \hline
L =&8&0.00990(3)&0.70(6)&0.86(2)&1.5(2)&\\ \hline
L =&10&0.00128(6)&0.79(5)&0.808(16)&2.0(6)&0.806(20)\\ \hline
\end{tabular}
\end{center}
\vspace{1cm}
\begin{center}     \Large\bf Table 2  \rm\normalsize
\end{center}
The interface tension $\sigma$ and the prefactor $C$ obtained from a fit
of the $L$-dependence of $E_{0a}$ according to (\ref{E0afit}).
\begin{center}
\begin{tabular}{|c||c|c|c|} \hline
$\beta$&$\sigma$&$\ln C$&$C$\\ \hline \hline
0.2275&0.0150(1)&-1.675(19)&0.187(4) \\ \hline
0.2327&0.0328(2)&-1.260(13)&0.284(4)\\ \hline
0.2391&0.0572(2)&-0.955(10)&0.385(4)\\ \hline
\end{tabular}
\end{center}
\newpage
\begin{center}     \Large\bf Table 3  \rm\normalsize
\end{center}
Renormalized parameters determined from the numerical simulation.
\begin{center}
\begin{tabular}{|c||*{5}{c|}} \hline
$\beta$& $m$& $m_R$& $v_R$& $Z_R$& $u_R$ \\
\hline \hline
0.2275& 0.416(16) &0.44(05) &0.262(30) &0.80(19) &19(7) \\ \hline
0.2327& 0.619(14) &0.62(12) &0.34(06) &0.71(27) &16(9) \\ \hline
0.2391& 0.806(20)  &0.80(32) &0.40(16) &0.65(52) &15(18) \\ \hline
\end{tabular}
\end{center}
\vspace{1cm}
\begin{center}     \Large\bf Table 4  \rm\normalsize
\end{center}
The dimensionless ratios $\sigma/m^2$ and $C/m$ as obtained in the
numerical simulation.
The columns to the right of these ratios contain the coupling $u_R$
calculated from the ratio according to the semiclassical expressions
(\ref{sigma}) and (\ref{C}), respectively.
\begin{center}
\begin{tabular}{|c||c|c||c|c|} \hline
$\beta$&$\sigma / m^2$&$u_R$&$C/m$&$u_R$ \\
\hline \hline
0.2275& 0.087(7) & 17.1(1.0) & 0.450(20) & 18.0(1.6) \\ \hline
0.2327& 0.086(4) & 17.2(0.6) & 0.458(12) & 17.4(0.9) \\ \hline
0.2391& 0.088(4) & 16.9(0.6) & 0.477(13) & 16.1(0.9) \\ \hline
\end{tabular}
\end{center}
\vspace{1cm}
\begin{center}     \Large\bf Figure Captions \rm\normalsize
\end{center}
{\bf Fig.~1}: The low-lying spectrum at $\beta = 0.2275$ as a function
of the lattice size $L$.\\[5mm]
{\bf Fig.~2}: The logarithm of the energy splitting $E_{0a}$ versus the
lattice cross-section $L^2$ for three values of the inverse temperature
$\beta$.
Also shown are the fits according to (\ref{E0afit}).\\[5mm]
{\bf Fig.~3}: The logarithm of the interface tension $\sigma$ versus
$\ln (1 - \beta_c / \beta)$.
Also shown is the fit according to the scaling law (\ref{sigscal}).

\newpage
\begin{thebibliography}{99}
%
\bibitem{AHK2}
K.\,Jansen, I.\,Montvay, G.\,M\"unster, T.\,Trappenberg, U.\,Wolff,\\
Nucl.\ Phys.\ \underline{B\,322} (1989) 698
%
\bibitem{Pegg}
I.\,A.\,McLure, I.\,L.\,Pegg,
J.\ Mol.\ Structure \underline{80} (1982) 393
%
\bibitem{Gielen}
H.\,L.\,Gielen, J.\ Thoen, O.\,B.\,Verbeke,
J.\ Chem.\ Phys.\ \underline{81} (1984) 6154
%
\bibitem{Moldover}
M.\,R.\,Moldover,
Phys.\ Rev.\ \underline{A\,31} (1985) 1022
%
\bibitem{Chaar}
H.\,Chaar, M.\,Moldover, J.\,Schmidt,
J.\ Chem.\ Phys.\ \underline{85} (1986) 418
%
\bibitem{Widom2}
B.\,Widom,
J.\ Chem.\ Phys.\ \underline{43} (1965) 3892
%
\bibitem{Widom1}
B.\,Widom,
in: Phase Transitions and Critical Phenomena,
Vol.2, ed.: C.\,Domb, M.\,S.\,Green,
Academic Press, London, 1971
%
\bibitem{Beysens}
D.\,Beysens,
in: Phase Transitions, Carg\`{e}se 1980, ed.: M.\,L\'{e}vy et al.,
Plenum Press, New York, 1982;\\
D.\,Beysens, A.\,Bourgou, P.\,Calmettes,
Phys.\ Rev.\ \underline{A\,26} (1982) 3589
%
\bibitem{Sengers}
J.\,V.\,Sengers
in: Phase Transitions, Carg\`{e}se 1980, ed.: M.\,L\'{e}vy et al.,
Plenum Press, New York, 1982
%
\bibitem{Kumar}
A.\,Kumar et al.,
Phys.\ Reports \underline{C\,98} (1983) 57
%
\bibitem{Zinn}
J.-C.\,Le Guillou, J.\,Zinn-Justin,
Phys.\ Rev.\ \underline{B\,21} (1980) 3976;\\
J.\,Zinn-Justin,
in: Phase Transitions, Carg\`{e}se 1980, ed.: M.\,L\'{e}vy et al.,
Plenum Press, New York, 1982
%
\bibitem{Fisk}
S.\,Fisk, B.\,Widom,
J.\ Chem.\ Phys.\ \underline{50} (1969) 3219
%
\bibitem{Stauffer}
D.\,Stauffer, M.\,Ferer, M.\,Wortis,
Phys.\ Rev.\ Letters \underline{29} (1972) 345
%
\bibitem{Pant}
B.\,B.\,Pant,
Ph.D.\ thesis, University of Pittsburgh, 1983
%
\bibitem{BF}
E.\,Br\'{e}zin, S.\,Feng,
Phys.\ Rev.\ \underline{B\,29} (1984) 472
%
\bibitem{Mue3d}
G.\,M\"unster,
Nucl.\ Phys.\ \underline{B\,340} (1990) 559
%
\bibitem{Binder}
K.\,Binder,
Phys.\ Rev.\ \underline{A\,25} (1982) 1699
%
\bibitem{Mon}
K.\,K.\,Mon,
Phys.\ Rev.\ Letters \underline{60} (1988) 2749
%
\bibitem{Meyer}
H.\,Meyer-Ortmanns, T.\,Trappenberg,
J.\ Stat.\ Phys.\ \underline{58} (1990) 185;\\
H.\,Meyer-Ortmanns,
Proceedings of the XXIII.\ International Symposium on the Theory of
Elementary Particles,
ed.: E.\,Wieczorek, Ahrenshoop, GDR, 1989
%
\bibitem{JanShen}
K.\,Jansen, Y.\,Shen,
San Diego preprint UCSD/PTH 92-02, BNL-45082, Jan.~1992
%
\bibitem{AHK1}
K.\,Jansen, J.\,Jers\'ak, I.\,Montvay, G.\,M\"unster,
T.\,Trappenberg, U.\,Wolff,\\
Phys.\ Letters \underline{B\,213} (1988) 203
%
\bibitem{BoIm}
C.\,Borgs, J.\,Z.\,Imbrie,
Harvard preprint 91-0233
%
\bibitem{Rough}
H.\,van Beijeren, I.\,Nolden,
The roughening transition,
in {\it Topics in Current Physics, Vol.~43: Structure and Dynamics of
Surfaces II},
ed.\ W.\,Schommers and P.\,von Blanckenhagen
(Springer, Berlin, 1987).
%
\bibitem{MaGo}
M.\,G\"opfert, G.\,Mack,
Comm.\ Math.\ Phys.\ \underline{82} (1981) 545
%
\bibitem{Adler}
J.\,Adler,
J.\ Phys.\ \underline{A\,16} (1983) 3585
%
\bibitem{Pawl}
G.\,S.\,Pawley et al.,
Phys.\ Rev.\ \underline{B\,29} (1984) 4030
%
\bibitem{Domb}
C.\,Domb,
Adv.\ Phys.\ \underline{9} (1960) 149
%
\bibitem{Fisher}
M.\,E.\,Fisher,
J.\ Phys.\ Soc.\ Japan Suppl.\ \underline{26} (1969) 87
%
\bibitem{PF}
V.\,Privman, M.\,E.\,Fisher,
J.\ Stat.\ Phys.\ \underline{33} (1983) 385
%
\bibitem{Mue4d}
G.\,M\"unster,
Nucl.\ Phys.\ \underline{B\,324} (1989) 630
%
\bibitem{SWA}
R.\,H.\,Swendsen, J.-S.\,Wang,
Phys.\ Rev.\ Letters \underline{58} (1987) 86
%
\bibitem{Wolff}
U.\,Wolff,
Phys.\ Rev.\ Letters \underline{62} (1989) 361
%
\bibitem{TaFi}
H.\,B.\,Tarko, M.\,E.\,Fisher,
Phys.\ Rev.\ \underline{B\,11} (1975) 1217
%
\end{thebibliography}
\end{document}